\documentclass[12pt]{article}
\usepackage{tikz}
\usetikzlibrary{matrix,arrows,decorations.pathmorphing}
\usepackage{jheppub}
\usepackage{amsmath,amssymb,euscript,array,mathrsfs}
\usepackage{arydshln}
\usepackage{todonotes}
\usepackage{graphicx}

\usepackage{epsfig}

\def\XXint#1#2#3{{\setbox0=\hbox{$#1{#2#3}{\int}$}
     \vcenter{\hbox{$#2#3$}}\kern-.5\wd0}}

\newcommand{\IM}{\operatorname{Im}}

\newcommand{\RE}{\operatorname{Re}}
\def\AMP{{\EuScript A}}

\def\d{\delta}

\def\Dbarslash{\,\,{\raise.15ex\hbox{/}\mkern-12mu {\bar D}}}
\def\Dslash{\,\,{\raise.15ex\hbox{/}\mkern-12mu D}}
\def\delslash{\,\,{\raise.15ex\hbox{/}\mkern-9mu \partial}}
\def\delbarslash{\,\,{\raise.15ex\hbox{/}\mkern-9mu {\bar\partial}}}

\newcommand{\EQ}[1]{\begin{equation*}\begin{split} #1 \end{split}\end{equation*}}

\title{\centerline{The Unbearable Beingness of Light} 
\vskip0.3cm
\centerline{\large -- Dressing and Undressing Photons in Black Hole Spacetimes --}}
\author{Timothy J. Hollowood and Graham M. Shore}
\affiliation{Department of Physics, Swansea University,
Swansea, SA2 8PP, UK.}
\emailAdd{t.hollowood@swansea.ac.uk, g.m.shore@swansea.ac.uk}

\abstract{Gravitational tidal forces acting on the virtual $e^+ e^-$
  cloud surrounding a photon endow spacetime with a non-trivial
  refractive index. This has remarkable properties unique to
  gravitational theories including superluminal low-frequency
  propagation, in apparent violation of causality, and amplification
  of the renormalized photon field, in apparent violation of
  unitarity. Using the geometry of null congruences and the Penrose
  limit, we illustrate these phenomena and their resolution by tracing
  the history of a photon as it falls into the near-singularity region
  of a black hole.

\vspace{1cm}
\begin{center}
{\bf Essay awarded third prize in the Gravity Research Foundation essay competition 2012}  \end{center}

}

\notoc
\begin{document}
\maketitle

\newpage

\noindent 1.~ In general relativity, photons propagate along null
geodesics, tracing out the causal structure of the background
spacetime. Along these null paths, the proper time vanishes---in 
their frame, photons exist only in an instant. They have no history; 
no past, no future, no time in which to evolve or decay. They are
caught forever in a single instant of time---the unbearable 
``beingness" of light \cite{Kundera}.

Quantum field theory, however, profoundly changes this
picture. Vacuum polarization envelops the photon in a cloud of virtual
$e^+e^-$ pairs. This dressed photon acquires an effective size, given
by the Compton wavelength $\lambda_c = h/mc$ of the electron, and
propagates as if it were an extended object, subject to gravitational
tidal forces. This changes the nature of photon propagation as the
phase velocity responds to the gravitational field and becomes
frequency-dependent. In QED, photons see curved spacetime as an
optical medium with a non-trivial refractive index.

The propagation of light in curved spacetime therefore exhibits all
the usual optical phenomena associated with a refractive index
$n(\omega)$, including dispersion and dissipation. The size of these 
effects is ${\cal O}(\alpha (\lambda_c/L)^2)$, where $L$ is a typical
curvature scale and $\alpha$ is the fine structure constant. 
However, there are surprises -- as first discovered by Drummond and
Hathrell \cite{Drummond:1979pp}, 
the low-frequency limit of the phase velocity 
can be superluminal, in apparent violation of causality. Moreover, the 
imaginary part of the refractive index can be
negative \cite{Hollowood:2008kq}, indicating 
an amplification rather than dissipation of light as it passes through 
curved spacetime, in apparent violation of unitarity. 

Reconciling these phenomena with the fundamental principles of quantum
field theory and general relativity has been the focus of our
investigations of this subject \cite{Hollowood:2007kt,Hollowood:2007ku,
Hollowood:2008kq,Hollowood:2009qz,Hollowood:2010bd,
Hollowood:2011yh}.
We find that indeed causality and
unitarity are respected, but at the expense of radical changes to many
well-established principles of QFT in flat spacetime, notably the
optical theorem and the range of theorems associated with analyticity,
including the vital Kramers-Kronig dispersion relation. In particular,
we find that the refractive index, now position-dependent, is governed
by the whole past trajectory of the photon. In QFT, photons are
liberated; they have a history, and a rich and fascinating one.

In this essay, we illustrate all these phenomena by following the
history of a photon as it falls into a black hole, showing how the
superluminal phase velocity is reconciled with causality and
describing the effects of the gravitational tidal forces on the
virtual $e^+e^-$ cloud. Remarkably, we find that the photon can 
become undressed---the tidal forces squeeze the vacuum 
polarization cloud, reducing the screening and amplifying
the renormalized quantum field amplitude as the photon 
approaches the singularity. This, then, is the photon's tale.

\begin{figure}[h]
\begin{center}
\begin{tikzpicture}[scale=1]
\draw[color=blue,very thick] (1,0) arc (0:90:1);
\draw[->,color=blue,very thick] (-1,0) arc (180:90:1);
\draw[->,color=red,very thick] (-1,0) arc (-180:-90:1);
\draw[color=red,very thick] (1,0) arc (0:-90:1);
\node[color=blue] at (-1.2,1) (i1) {$e^+$};
\node[color=red] at (1.2,-1) (i2) {$e^-$};
\node at (-2.5,0) (i3) {$\gamma$};
\node at (2.5,0) (i3) {$\gamma$};
\draw[decorate,decoration={snake,amplitude=0.1cm},very thick] (-2,0) -- (-1,0); 
\draw[decorate,decoration={snake,amplitude=0.1cm},very thick] (2,0) -- (1,0); 
\filldraw[black] (-1,0) circle (2pt);
\filldraw[black] (1,0) circle (2pt);
\node[rotate=45,opacity=0.2,scale=2.2] {};
% {curvature};
\end{tikzpicture}  
\end{center}
\caption{\small The one-loop Feynman diagram contributing to the
  vacuum polarization in QED.}
\label{figFeynman}
\end{figure}
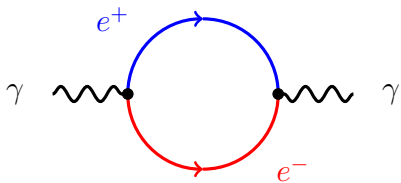

\noindent 2.~  The most illuminating way to describe these effects 
in QED is in terms of an initial value problem for the photon field 
$A_\mu(x)$, which satisfies a wave equation incorporating the one-loop
vacuum polarization given by the Feynman diagram in 
Fig.~\ref{figFeynman}. In the eikonal approximation, the
solution is
\EQ{
 A_\mu^{(i)}(x) = g(u)^{-{1/4}} 
{\cal A}(u_0) e^{i\omega V} e^{i\omega \int_{u_0}^u du'
(n(u';\omega) - 1)_{ij}}   \varepsilon_\mu^{(j)}(u) \ ,
}
where $u$ is a null coordinate along the classical light ray $\gamma$,
and $V$ is the corresponding null coordinate labelling geodesics 
in the null congruence around $\gamma$.
 ${\cal A}(u_0)$ is the amplitude on the initial value surface 
$u=u_0$, $\varepsilon_\mu^{(i)}(u)$ is the polarization
vector and  $n_{ij}(u;\omega)$  is a position, polarization and 
frequency-dependent refractive index.

A key observation is that to leading order in $(\lambda_c/L)^2$, 
we can replace the background spacetime by its Penrose limit 
\cite{Penrose} associated with the null geodesic $\gamma$. 
We have shown, using both worldline and conventional QFT methods 
\cite{Hollowood:2007kt,
Hollowood:2007ku,Hollowood:2008kq,Hollowood:2009qz},
how the refractive index is governed by the geometry of geodesic 
deviation around $\gamma$. 
But this is precisely the property of the spacetime which is
encoded in the Penrose limit. This can be seen by writing the metric
around $\gamma$ in null Fermi coordinates and making the 
Penrose rescaling $u\rightarrow u, v \rightarrow \lambda^2 v, 
x^i \rightarrow \lambda x^i$; truncating at ${\cal O}(\lambda^2)$ then
recovers the Brinkmann plane wave metric 
$ds^2 = 2 du dv - h_{ij}(u) x^i x^j du^2 + dx^i dx^i$,
with the profile function identified 
as $h_{ij}= R_{uiuj}$ \cite{Blau:2006ar}. 
Since the geodesic deviation equation is written 
in terms of the Jacobi fields $x^i$ as $d^2x^i/du^2 = R^i{}_{uju}x^j$, this
isolates exactly the required curvature components.
In QED, the expansion in $\lambda$ is mirrored as an expansion
of the Green functions in $\lambda_c/L$, which ensures that 
the gravitational curvature is relatively weak on the quantum scale.

\begin{figure}[h]
  \begin{minipage}[t]{.47\textwidth}
 \EQ{
&\frac{d^2x^i}{du^2} = R^i{}_{uju}x^j\ ,\\[7pt]
&x^i(u) = A_{ij}(u,u') \frac{dx^j(u')}{du'} \ ,\\[7pt]
&A_{ij}(u',u') = 0,\quad \frac{dA_{ij}(u,u')}{du}\Big|_{u=u'} = \delta_{ij}\ .
}
\end{minipage}
  \hfill
  \begin{minipage}[t]{.47\textwidth}
    \begin{center}\begin{tikzpicture}[scale=0.8]
\draw[->] (-0.5,-1) to[out=70,in=240] (0,0) to[out=60,in=200] (4,3) to[out=20,in=200] (5,3.3);
\draw[->,color=red]  (0,0) to[out=80,in=190] (4,3) to [out=10,in=170] (5,3);
\draw[->,color=red]  (0,0) to[out=20,in=250] (4,3) to [out=70,in=275] (4.1,4);
\draw[->,color=blue] (0,0) to[out=60,in=270] (1,4);
\draw[->,color=blue] (0,0) to[out=50,in=140] (4,-0.5);
\filldraw (0,0) circle (2pt);
\filldraw (4,3) circle (2pt);
\node at (-0.7,-1.3) (i1) {$\gamma$};
\node at (0.5,-0.2) (i2) {$u'$};
\node at (5.5,2) (i3) {conjugate point};
\draw[->] (i3) -- (4.2,2.8);
\node at (-1,3) (i4) {diverging};
\draw[->] (i4) -- (0.8,2.5);
\node at (4.5,1) (i5) {converging};
\draw[->] (i5) -- (3.4,1.8);
\draw[->] (-0.6,-0.5) -- (1.6,-0.8);
\draw[->] (-0.3,-0.8) -- (-0.3,1.4);
\node at (2,-0.9) (j1) {$x^1$};
\node at (-0.8,1.2) (j2) {$x^2$};
\end{tikzpicture}
    \end{center}
  \end{minipage}
\caption{\small Null geodesic spray from a point $u'$ on $\gamma$. On the
  left is the equation for geodesic deviation of the Jacobi fields
  $x^i(u)$ with a solution written in terms of the $2\times 2$ matrix
  $A_{ij}(u,u')$. The boundary conditions on the spray are written in the last line. 
  In the illustration on the right, geodesics in the spray at $u=u'$ diverge
  in one direction (blue) and converge in the other (red) to cross at a conjugate point.}
\label{f13}
\end{figure}
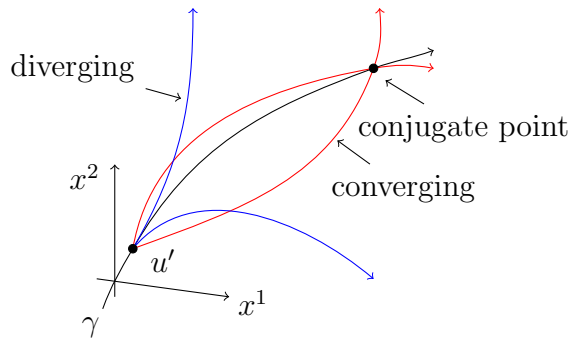

The refractive index is expressed entirely in terms of
geometric quantities related to geodesic deviation. Solving the Jacobi
equation with ``geodesic spray" boundary conditions at $u=u'$, 
as in Fig.~\ref{f13},
determines the Van
Vleck-Morette matrix $\Delta_{ij}(u,u') = (u'-u) A^{-1}_{ji}(u,u')$.
We have then shown that the refractive index of curved spacetime 
in QED with scalar electrons, with initial value surface $u_0\rightarrow -\infty$, is 
\begin{equation*}
n_{ij}(u;\omega) = \d_{ij} - {i\alpha\over{2\pi\omega}}
\int_0^1 d\xi\, \xi(1-\xi)  \int_{-\infty}^u {du'\over{(u-u')^2}} 
e^{-{im^2(u-u')\over{2\omega\xi(1-\xi)}}} ~\Big[
\Delta_{ij}(u,u')\sqrt{\Delta(u,u')} - \d_{ij}\Big] \ .
\end{equation*}
(A similar, more complicated, formula holds for spinor electrons.)
Notice the key point that $n_{ij}(u,\omega)$ is given by an integral
over the whole of the past worldline $u'<u$ of the photon. 
In curved spacetime, photons have a history, their local propagation 
characteristics being determined by the curvature throughout their path.

The refractive index has novel analyticity properties, related to the
existence of conjugate points of the null geodesic congruence.
This has far-reaching consequences 
\cite{Hollowood:2008kq} since it shows that
amplitudes in curved spacetime have geometric as well as kinematic 
branch points and cuts, radically modifying many apparently
fundamental theorems in quantum field theory and S-matrix theory.
In particular, the Kramers-Kronig
dispersion relation in curved spacetime becomes simply
\begin{equation*}
 n(u;\infty) = n(u;0) - {1\over{i\pi}} {\cal P} \int_{-\infty}^\infty 
{d\omega\over\omega}\, n(u,\omega) \ .
\end{equation*}
In flat spacetime, hermitian analyticity would then imply that 
the principal value integral can be written in terms of $\IM
n(u;\omega)$, which would be guaranteed by the optical theorem to be
positive, forcing $n(\infty) < n(0)$.  However, causality depends on 
the wavefront velocity (which is identified with the high frequency limit 
of the phase velocity \cite{Shore:2003jx}) not exceeding $c$,
that is $n(\infty) \geq 1$. In flat spacetime,
this would clearly be incompatible with a
superluminal low-frequency value $n(0) < 1$. In curved spacetime,
however, hermitian analyticity is violated due to the geometric cuts
in the complex $\omega$-plane in $n(u;\omega)$ and the 
conventional Kramers-Kronig relation fails. 

Moreover, we also find instances where $\IM n(u;\omega)$ 
is negative, requiring a reappraisal of the optical theorem 
itself \cite{Hollowood:2010bd,Hollowood:2011yh}.
The key point is that in curved spacetime, the optical theorem, which
relates the imaginary part of scattering amplitudes to cross sections,
only holds globally. For example, here we can express the total
probability up to time $u$ for the decay $\gamma\rightarrow e^+ e^-$ as
\begin{equation*}
P_{\gamma\rightarrow e^+ e^-}(u) = 4\omega \int_{u_0}^u du^{\prime}  \IM
n(u';\omega)  \ .
\end{equation*}
which is manifestly positive. However, without $u$-translation
invariance, the local version is no longer necessarily positive and 
does not represent a true decay rate, removing the apparent conflict 
between $\IM n(u;\omega) < 0$ and unitarity. The field may be locally
amplified, but integrated over its whole past trajectory there must be a
net attenuation.

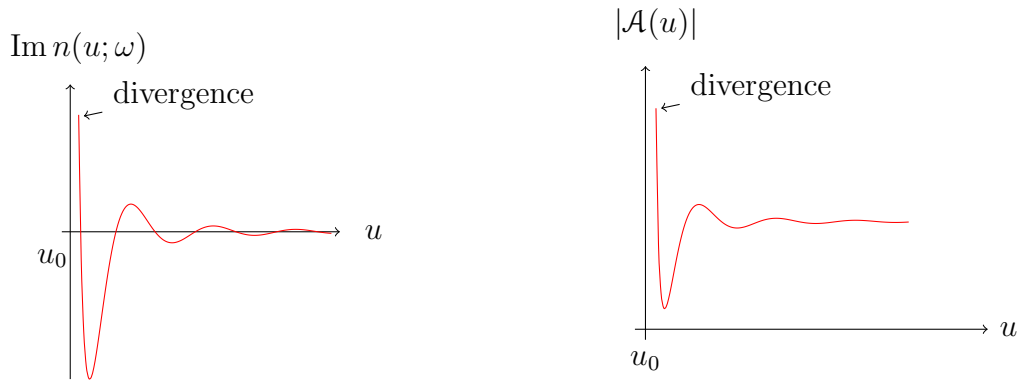
\begin{figure}[h]
  \begin{minipage}[t]{.47\textwidth}
    \begin{center}
\begin{tikzpicture}[scale=0.7]
\begin{scope}[xscale=0.8,yscale=14]
\draw[->] (-0.4,0) -- (6.2,0);
\draw[->] (-0.20,-0.2) --(-0.2,0.2);
\draw[color=red] plot[smooth] coordinates {(0., 0.158721)  (0.06, -0.0363973)  (0.12, -0.135358)  (0.18,  
-0.182326)  (0.24, -0.19898)  (0.3, -0.197279)  (0.36, -0.184393)   
(0.42, -0.164896)  (0.48, -0.141837)  (0.54, -0.117313)  (0.6,  
-0.0927877)  (0.66, -0.0692897)  (0.72, -0.0475299)  (0.78,  
-0.0279819)  (0.84, -0.0109374)  (0.9, 0.00345546)  (0.96, 
  0.0151615)  (1.02, 0.0242358)  (1.08, 0.0308045)  (1.14, 
  0.035049)  (1.2, 0.0371919)  (1.26, 0.037485)  (1.32, 
  0.0361985)  (1.38, 0.0336116)  (1.44, 0.0300039)  (1.5, 
  0.0256479)  (1.56, 0.0208032)  (1.62, 0.0157104)  (1.68, 
  0.0105874)  (1.74, 0.00562585)  (1.8, 
  0.00098868)  (1.86, -0.00319111)  (1.92, -0.00681106)  (1.98,  
-0.00979896)  (2.04, -0.0121117)  (2.1, -0.0137336)  (2.16,  
-0.0146742)  (2.22, -0.0149651)  (2.28, -0.0146571)  (2.34,  
-0.0138168)  (2.4, -0.0125228)  (2.46, -0.0108623)  (2.52,  
-0.00892733)  (2.58, -0.00681152)  (2.64, -0.00460678)  (2.7,  
-0.00240051)  (2.76, -0.000273038)  (2.82, 0.00170443)  (2.88, 
  0.00347147)  (2.94, 0.00497965)  (3., 0.0061932)  (3.06, 
  0.00708933)  (3.12, 0.00765801)  (3.18, 0.00790144)  (3.24, 
  0.00783307)  (3.3, 0.00747643)  (3.36, 0.0068636)  (3.42, 
  0.00603358)  (3.48, 0.00503049)  (3.54, 0.00390173)  (3.6, 
  0.00269618)  (3.66, 0.00146236)  (3.72, 
  0.000246851)  (3.78, -0.000907227)  (3.84, -0.0019615)  (3.9,  
-0.00288338)  (3.96, -0.00364686)  (4.02, -0.00423306)  (4.08,  
-0.00463046)  (4.14, -0.0048349)  (4.2, -0.00484933)  (4.26,  
-0.00468331)  (4.32, -0.00435236)  (4.38, -0.00387705)  (4.44,  
-0.00328206)  (4.5, -0.00259508)  (4.56, -0.00184573)  (4.62,  
-0.00106434)  (4.68, -0.00028094)  (4.74, 0.000475847)  (4.8, 
  0.00117972)  (4.86, 0.00180751)  (4.92, 0.00233988)  (4.98, 
  0.00276181)  (5.04, 0.00306291)  (5.1, 0.00323764)  (5.16, 
  0.00328522)  (5.22, 0.00320948)  (5.28, 0.00301854)  (5.34, 
  0.00272429)  (5.4, 0.00234185)  (5.46, 0.00188886)  (5.52, 
  0.00138476)  (5.58, 0.000850058)  (5.64, 
  0.000305487)  (5.7, -0.000228686)  (5.76, -0.00073339)  (5.82,  
-0.00119137)  (5.88, -0.00158776)  (5.94, -0.00191046)  (6.,  
-0.00215052)};
\end{scope}
\node at (0,3.5) (i1) {$\IM n(u;\omega)$};
\node at (5.6,0) (i2) {$u$};
\node at (-0.5,-0.5) (i3) {$u_0$};
\node at (2,2.6) (i6) {divergence};
\draw[->] (i6) -- (0.1,2.2);
\end{tikzpicture}
    \end{center}
  \end{minipage}
  \hfill
  \begin{minipage}[t]{.47\textwidth}
    \begin{center}
\begin{tikzpicture}[scale=0.7]
\begin{scope}[xscale=0.8,yscale=16,yshift=-0.8cm]
\draw[color=red] plot[smooth] coordinates {(0., 1.)  (0.06, 0.837345)  (0.12, 0.780695)  (0.18, 
  0.763216)  (0.24, 0.763985)  (0.3, 0.7738)  (0.36, 0.78789)  (0.42, 
  0.803546)  (0.48, 0.81916)  (0.54, 0.833763)  (0.6, 
  0.846792)  (0.66, 0.857947)  (0.72, 0.86711)  (0.78, 
  0.874284)  (0.84, 0.879562)  (0.9, 0.883092)  (0.96, 
  0.885065)  (1.02, 0.885691)  (1.08, 0.885194)  (1.14, 
  0.883797)  (1.2, 0.881718)  (1.26, 0.879162)  (1.32, 
  0.876317)  (1.38, 0.873353)  (1.44, 0.870416)  (1.5, 
  0.867629)  (1.56, 0.86509)  (1.62, 0.862875)  (1.68, 
  0.861035)  (1.74, 0.859602)  (1.8, 0.858586)  (1.86, 
  0.85798)  (1.92, 0.857763)  (1.98, 0.857901)  (2.04, 
  0.858349)  (2.1, 0.859057)  (2.16, 0.859969)  (2.22, 
  0.861027)  (2.28, 0.862176)  (2.34, 0.863358)  (2.4, 
  0.864524)  (2.46, 0.865628)  (2.52, 0.866631)  (2.58, 
  0.867502)  (2.64, 0.868217)  (2.7, 0.868761)  (2.76, 
  0.869126)  (2.82, 0.869311)  (2.88, 0.869324)  (2.94, 
  0.869178)  (3., 0.868889)  (3.06, 0.868481)  (3.12, 
  0.867978)  (3.18, 0.867407)  (3.24, 0.866795)  (3.3, 
  0.86617)  (3.36, 0.865556)  (3.42, 0.864977)  (3.48, 
  0.864454)  (3.54, 0.864004)  (3.6, 0.86364)  (3.66, 0.86337)  (3.72,
   0.863199)  (3.78, 0.863129)  (3.84, 0.863155)  (3.9, 
  0.863272)  (3.96, 0.863468)  (4.02, 0.863733)  (4.08, 
  0.864051)  (4.14, 0.864407)  (4.2, 0.864785)  (4.26, 
  0.86517)  (4.32, 0.865545)  (4.38, 0.865898)  (4.44, 
  0.866214)  (4.5, 0.866483)  (4.56, 0.866699)  (4.62, 
  0.866854)  (4.68, 0.866945)  (4.74, 0.866973)  (4.8, 
  0.866939)  (4.86, 0.866848)  (4.92, 0.866706)  (4.98, 
  0.86652)  (5.04, 0.8663)  (5.1, 0.866057)  (5.16, 0.8658)  (5.22, 
  0.865541)  (5.28, 0.865289)  (5.34, 0.865054)  (5.4, 
  0.864844)  (5.46, 0.864667)  (5.52, 0.864528)  (5.58, 
  0.86443)  (5.64, 0.864377)  (5.7, 0.864368)  (5.76, 
  0.864401)  (5.82, 0.864474)  (5.88, 0.864582)  (5.94, 0.86472)  (6.,
   0.864881)};
\end{scope}
\draw[->] (-0.4,-1) -- (6.3,-1);
\draw[->] (-0.2,-1.2) -- (-0.2,4);
\node at (0,4.8) (i1) {$|\AMP(u)|$};
\node at (6.7,-1) (i2) {$u$};
\node at (-0.2,-1.6) (i3) {$u_0$};
\node at (2,3.6) (i6) {divergence};
\draw[->] (i6) -- (0.1,3.2);
\end{tikzpicture}
   \end{center}
  \end{minipage}
  \caption{\small $\IM n(u;\omega)$ and the amplitude $|{\cal A}(u)|$ 
    as a function of $u$ in the near-flat spacetime region far from the
    black hole.  This shows the characteristic oscillatory transient 
    behaviour as the photon becomes dressed with its virtual 
    $e^+ e^-$ cloud, screening the bare field and reducing the
    amplitude. In QED the initial value of the field is tuned
    to be divergent to give a finite ${\cal A}(u)$ away from the 
    short-distance region close to $u_0$. }
\label{figQEDFlat}
\end{figure}

\noindent 3.~ We are now ready to follow the evolution of the photon
field as it falls towards the singularity of a black hole. Starting
from an initial value surface $u_0$ far from the hole where spacetime
is essentially flat, the imaginary part of the refractive index 
$\IM n(u;\omega)$ and the field amplitude 
${\cal A}(u)$, for either polarization, have the form shown in 
Fig.~\ref{figQEDFlat}.  

Now consider a planar null geodesic in Schwarzschild spacetime with
impact parameter $b$. In ref.~\cite{Hollowood:2009qz}, we developed new
techniques for determining the Penrose limit for null geodesics in a
general class of black hole spacetimes. We showed that the 
eigenvalues of the plane wave profile function $h_{ij}$  are
\begin{equation*}
h_{\pm} = {1\over 2} R_{\mu\nu}\hat k^\mu \hat k^\nu 
\pm {3\over 2} |\Psi_2|^{5/3} |K_s|^2  \ ,
\end{equation*}
where $\hat k^\mu$ is the tangent vector to $\gamma$,
$\Psi_2 = - C_{\mu\nu\lambda\rho}\ell^\mu m^\nu \bar m^\rho
n^\rho=-M/r^3$ is the Weyl tensor in the Newman-Penrose basis and $K_s$, 
with $|K_s|^2 = 2 M^{-{2/3}} b^2$, is the Walker-Penrose
conserved quantity \cite{Walker,Chandra} associated with the null
orbits. The Penrose limit
therefore has $h_{11} = -h_{22} =  3Mb^2/r^5$. 

An important simplification arises in the near-singularity limit, with
$h_{11} = -h_{22} = 6/25u^2$, 
independent of $M$ and $b$.
This is a singular homogeneous plane wave. Other Penrose limits are 
readily calculated, e.g.~the near-singularity limit of equatorial null orbits
in the Kerr black hole are also homogeneous plane waves with identical
$h_{ij}$, which is determined purely by the power-law nature of the
singularity \cite{Szekeres:1993qe,Celerier:2002yi,Blau:2004yi}.

\begin{figure}[h] 
\begin{center}
\begin{tikzpicture}[scale=1,fill=blue!20]
\node at (-7,-0.5) (q1) {$A_{11}(u,u')=5(uu')^{2/5}\Big(u^{1/5}-u^{\prime \,1/5}\Big)$};
\node at (-7,1.9) (q2) {$A_{22}(u,u')=\frac57(uu')^{-1/5}\Big(u^{7/5}-u^{\prime \,7/5}\Big)$};
\draw[fill] (-0.5,2.5) -- (3,2.5) -- (3,-1.5) -- (-0.5,-0.5) --(-0.5,2.5);
\filldraw[black] (1,0.7) circle (2pt);
\draw[-] (-2.5,-0.4) to[out=20,in=200] (1,0.7);
\draw[-] (-2.5,-0.4) to[out=200,in=20] (-2.9,-0.6);
\filldraw[black] (-2.5,-0.4) circle (2pt);
\draw[->,color=red] (-2.5,-0.4) .. controls (-1,-0.1) and (3.1,0.4) ..  (1,0.7);
\draw[->,color=red] (-2.5,-0.4) .. controls (-1,0.6) and (0,0.8) ..  (1,0.7);
\draw[->,color=blue] (-2.5,-0.4) .. controls (-1,-0.1) and (0.8,0.4) ..  (0.8,-1);
\draw[->,color=blue] (-2.5,-0.4) .. controls (-1,0.6) and (0.1,0.8) ..  (0.6,2.7);
\node[rotate=-45,opacity=0.2,scale=2.2] at (1.25,0.7) {singularity};
\node[color=blue] at (-2,1.8) (a1) {diverging};
\node[color=red] at (-1.5,-1.2) (a2) {converging};
\draw[->,color=red] (a2) -- (0.8,0.1);
\draw[->,color=red] (a2) -- (-3.8,-0.7);
\draw[->,color=blue] (a1) -- (-0.3,1.3);
\draw[->,color=blue] (a1) -- (-3.6,1.9);
\end{tikzpicture}
\end{center}
\caption{\small The behaviour of the null geodesic spray as the
  singularity is approached. In one plane the geodesics diverge (blue) while in the orthogonal plane geodesics 
  converge with the singularity being a conjugate point (red).}
\label{f10}
\end{figure}
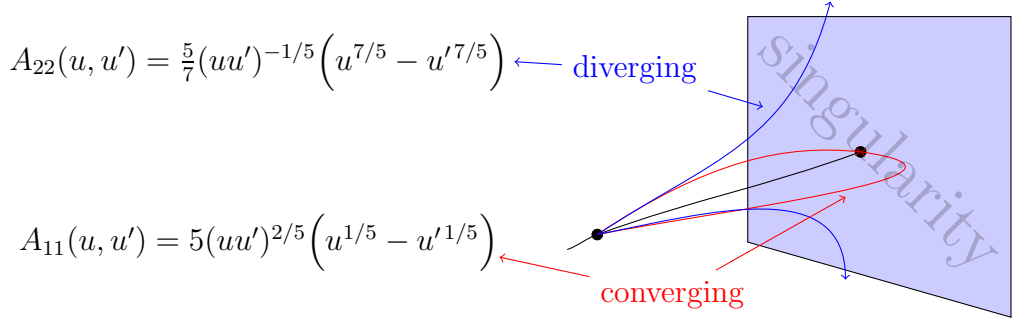

The refractive index in the near-singularity region now follows from
the geodesic spray matrix found from the Jacobi equation with 
this $h_{ij}$, illustrated in Fig.~\ref{f10}.

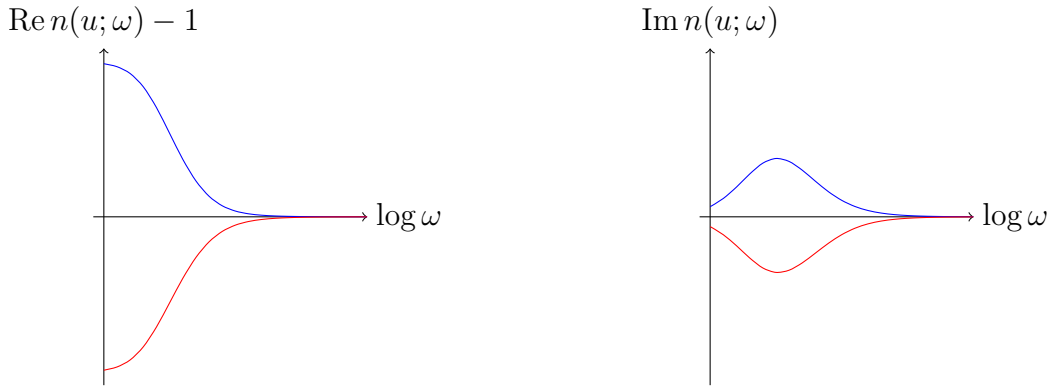
\begin{figure}[h]
  \begin{minipage}[t]{.47\textwidth}
    \begin{center}
\begin{tikzpicture}[scale=0.7,domain=0:5]
\draw[->] (-0.2,0) -- (5,0);
\draw[->] (0,-3.2) --(0,3.2);
\node at (0,3.7) (i1) {$\RE n(u;\omega)-1$};
\node at (5.8,0) (i2) {$\log\omega$};
\draw[color=blue] plot[smooth] coordinates {(0., 2.90966)  (0.25, 2.85337)  (0.5, 2.72334)  (0.75, 2.47058)  (1.,
   2.07555)  (1.25, 1.58435)  (1.5, 1.09119)  (1.75, 0.680542)  (2., 
  0.389114)  (2.25, 0.208887)  (2.5, 0.10889)  (2.75, 0.0568462)  (3.,
   0.0302459)  (3.25, 0.0164697)  (3.5, 0.00914749)  (3.75, 
  0.00515375)  (4., 0.00293105)  (4.25, 0.00167673)  (4.5, 
  0.000962569)  (4.75, 0.000553741)  (5., 0.000318942)};
\draw[color=red] plot[smooth] coordinates {(0., -2.91525)  (0.25, -2.85415)  (0.5, -2.72001)  (0.75, -2.4704)   
(1., -2.0933)  (1.25, -1.63257)  (1.5, -1.16814)  (1.75, -0.770617)   
(2., -0.473907)  (2.25, -0.27682)  (2.5, -0.157297)  (2.75,  
-0.088677)  (3., -0.0501172)  (3.25, -0.0284826)  (3.5, -0.0162706)   
(3.75, -0.0093294)  (4., -0.00536256)  (4.25, -0.00308711)  (4.5,  
-0.0017788)  (4.75, -0.00102551)  (5., -0.00059141)};
\end{tikzpicture}
    \end{center}
  \end{minipage}
  \hfill
  \begin{minipage}[t]{.47\textwidth}
    \begin{center}
\begin{tikzpicture}[scale=0.7]
\draw[->] (-0.2,0) -- (5,0);
\draw[->] (0,-3.2) --(0,3.2);
\node at (0,3.7) (i1) {$\IM n(u;\omega)$};
\node at (5.8,0) (i2) {$\log\omega$};
\draw[color=blue] plot[smooth] coordinates {(0., 0.191939)  (0.25, 0.348607)  (0.5, 0.564961)  (0.75, 
  0.812486)  (1., 1.01967)  (1.25, 1.10883)  (1.5, 1.05227)  (1.75, 
  0.886246)  (2., 0.675688)  (2.25, 0.47485)  (2.5, 0.312823)  (2.75, 
  0.196469)  (3., 0.119435)  (3.25, 0.0711037)  (3.5, 
  0.0417904)  (3.75, 0.0243737)  (4., 0.0141513)  (4.25, 
  0.00819448)  (4.5, 0.00473781)  (4.75, 0.00273682)  (5., 
  0.00158012)};
\draw[color=red] plot[smooth] coordinates {(0., -0.185434)  (0.25, -0.34043)  (0.5, -0.549115)  (0.75,  
-0.781162)  (1., -0.971713)  (1.25, -1.0565)  (1.5, -1.01466)  (1.75,  
-0.875275)  (2., -0.689125)  (2.25, -0.500514)  (2.5, -0.338814)   
(2.75, -0.216943)  (3., -0.133546)  (3.25, -0.0801225)  (3.5,  
-0.0473102)  (3.75, -0.0276687)  (4., -0.0160901)  (4.25,  
-0.0093258)  (4.5, -0.0053948)  (4.75, -0.0031173)  (5., -0.00180012)}
;\end{tikzpicture}
   \end{center}
  \end{minipage}
  \caption{\small $\RE n(u;\omega)$ and $\IM n(u;\omega)$ 
  as a function of frequency
  $\omega$ for fixed position $u$ in the near singularity region of a
  Schwarzschild black hole. The red curves denote the 
  polarization perpendicular to the orbital plane, corresponding to
  a converging null congruence, while the blue curves denote the
  polarization lying in the orbital plane, for which the null
  congruence is diverging. Note that $n(u;\omega)\rightarrow 1$
  as $\omega\rightarrow\infty$, as required to maintain causality. } 
\label{BHQEDnomega}
\end{figure}

The frequency dependence of $n(u;\omega)$ for fixed $u$ is shown in
Fig.~\ref{BHQEDnomega}. 
This shows explicitly how $\RE n(u;\omega) \rightarrow 1$
as $\omega\rightarrow \infty$ for both polarizations, including the 
polarization $\varepsilon^{(1)}$ which is superluminal at low
frequency, ensuring causality is maintained. The converging
polarization $\varepsilon^{(1)}$ also has $\IM n(u;\omega)$ negative,
which would violate the conventional optical theorem
but is consistent with our new curved spacetime formulation.

\begin{figure}[ht] 
\begin{center}
\begin{tikzpicture}[scale=0.6]
\begin{scope}[xscale=1,yscale=0.02]
\draw[color=red] plot[smooth] coordinates {(0., -12.3369)  (0.06, -12.4625)  (0.12, -12.5904)  (0.18,  
-12.7208)  (0.24, -12.8537)  (0.3, -12.9892)  (0.36, -13.1273)   
(0.42, -13.2682)  (0.48, -13.4119)  (0.54, -13.5585)  (0.6,  
-13.7081)  (0.66, -13.8608)  (0.72, -14.0166)  (0.78, -14.1757)   
(0.84, -14.3382)  (0.9, -14.5042)  (0.96, -14.6738)  (1.02, -14.847)   
(1.08, -15.0242)  (1.14, -15.2052)  (1.2, -15.3904)  (1.26,  
-15.5798)  (1.32, -15.7736)  (1.38, -15.9718)  (1.44, -16.1748)   
(1.5, -16.3826)  (1.56, -16.5955)  (1.62, -16.8135)  (1.68, -17.037)   
(1.74, -17.266)  (1.8, -17.5008)  (1.86, -17.7417)  (1.92, -17.9888)   
(1.98, -18.2424)  (2.04, -18.5028)  (2.1, -18.7701)  (2.16,  
-19.0448)  (2.22, -19.3271)  (2.28, -19.6173)  (2.34, -19.9158)   
(2.4, -20.2229)  (2.46, -20.539)  (2.52, -20.8644)  (2.58, -21.1996)   
(2.64, -21.5451)  (2.7, -21.9013)  (2.76, -22.2687)  (2.82,  
-22.6478)  (2.88, -23.0392)  (2.94, -23.4435)  (3., -23.8614)  (3.06,  
-24.2935)  (3.12, -24.7405)  (3.18, -25.2033)  (3.24, -25.6827)   
(3.3, -26.1795)  (3.36, -26.6947)  (3.42, -27.2295)  (3.48,  
-27.7847)  (3.54, -28.3618)  (3.6, -28.9619)  (3.66, -29.5865)   
(3.72, -30.237)  (3.78, -30.9152)  (3.84, -31.6227)  (3.9, -32.3615)   
(3.96, -33.1337)  (4.02, -33.9417)  (4.08, -34.7878)  (4.14,  
-35.6749)  (4.2, -36.6059)  (4.26, -37.5843)  (4.32, -38.6136)   
(4.38, -39.6979)  (4.44, -40.8417)  (4.5, -42.0499)  (4.56,  
-43.3282)  (4.62, -44.6827)  (4.68, -46.1204)  (4.74, -47.6492)   
(4.8, -49.2778)  (4.86, -51.0163)  (4.92, -52.8762)  (4.98,  
-54.8704)  (5.04, -57.014)  (5.1, -59.3242)  (5.16, -61.8211)  (5.22,  
-64.5282)  (5.28, -67.4727)  (5.34, -70.6873)  (5.4, -74.2105)   
(5.46, -78.0887)  (5.52, -82.3779)  (5.58, -87.1467)  (5.64,  
-92.4796)  (5.7, -98.4823)  (5.76, -105.289)  (5.82, -113.07)  (5.88,  
-122.052)  (5.94, -132.532)  (6., -144.917)};
\end{scope}
\begin{scope}[xscale=1,yscale=0.14]
\draw[color=blue] plot[smooth] coordinates {(0., 1.39294)  (0.06, 1.40766)  (0.12, 1.42268)  (0.18, 
  1.43802)  (0.24, 1.45367)  (0.3, 1.46965)  (0.36, 1.48597)  (0.42, 
  1.50263)  (0.48, 1.51966)  (0.54, 1.53706)  (0.6, 1.55485)  (0.66, 
  1.57303)  (0.72, 1.59162)  (0.78, 1.61064)  (0.84, 1.6301)  (0.9, 
  1.65002)  (0.96, 1.6704)  (1.02, 1.69127)  (1.08, 1.71265)  (1.14, 
  1.73456)  (1.2, 1.75701)  (1.26, 1.78002)  (1.32, 1.80361)  (1.38, 
  1.82782)  (1.44, 1.85266)  (1.5, 1.87815)  (1.56, 1.90433)  (1.62, 
  1.93122)  (1.68, 1.95884)  (1.74, 1.98724)  (1.8, 2.01645)  (1.86, 
  2.04649)  (1.92, 2.07741)  (1.98, 2.10924)  (2.04, 2.14202)  (2.1, 
  2.1758)  (2.16, 2.21063)  (2.22, 2.24654)  (2.28, 2.28361)  (2.34, 
  2.32187)  (2.4, 2.36139)  (2.46, 2.40224)  (2.52, 2.44447)  (2.58, 
  2.48817)  (2.64, 2.5334)  (2.7, 2.58026)  (2.76, 2.62883)  (2.82, 
  2.6792)  (2.88, 2.73147)  (2.94, 2.78577)  (3., 2.8422)  (3.06, 
  2.90089)  (3.12, 2.96199)  (3.18, 3.02564)  (3.24, 3.09201)  (3.3, 
  3.16127)  (3.36, 3.23361)  (3.42, 3.30926)  (3.48, 3.38843)  (3.54, 
  3.47138)  (3.6, 3.55838)  (3.66, 3.64975)  (3.72, 3.74581)  (3.78, 
  3.84694)  (3.84, 3.95354)  (3.9, 4.06608)  (3.96, 4.18506)  (4.02, 
  4.31106)  (4.08, 4.4447)  (4.14, 4.58671)  (4.2, 4.7379)  (4.26, 
  4.89918)  (4.32, 5.0716)  (4.38, 5.25635)  (4.44, 5.4548)  (4.5, 
  5.66854)  (4.56, 5.89939)  (4.62, 6.14951)  (4.68, 6.42139)  (4.74, 
  6.71801)  (4.8, 7.0429)  (4.86, 7.4003)  (4.92, 7.79535)  (4.98, 
  8.23431)  (5.04, 8.72493)  (5.1, 9.2769)  (5.16, 9.90248)  (5.22, 
  10.6174)  (5.28, 11.4424)  (5.34, 12.4048)  (5.4, 13.5422)  (5.46, 
  14.9069)  (5.52, 16.5747)  (5.58, 18.6591)  (5.64, 21.3385) };
  \end{scope}
\draw[->] (-0.2,0) -- (6.6,0);
\draw[->,decorate,decoration={snake,amplitude=0.05cm},very thick] (6.2,-3.4) --(6.2,3.4);
\node[color=blue,text width=3cm] at (1.3,3) (i6) {\small polarization 2: diverging};
\node[color=blue] at (8.4,2.3) (k1) {\small attenuation};
\draw[->,color=blue] (k1) -- (5.7,2.1);
\draw[->,color=blue] (i6) -- (3.2,0.7);
\node[color=red,text width=3.2cm] at (1,-2.2)  (i7) {\small polarization 1: converging};
\node[color=red] at (8.9,-2.3) (k2) {\small amplification};
\draw[->,color=red] (k2) -- (5.9,-2.1);
\draw[->,color=red] (i7) -- (4.4,-0.9);
\node at (6.4,3.7) (i1) {$\IM n(u)$};
\node at (6.9,0) (i2) {$u$};
\node at (8.7,-1) (i4) {\small singularity};
\draw[->] (i4) -- (6.4,-1.3);
\end{tikzpicture}
 \end{center}
\caption{\small $\IM n(u;\omega)$ 
  for the two polarizations in QED in a Schwarzschild
  black hole as $u$ approaches the singularity at $u=0$ from $u<0$.}
  \label{f11}
  \end{figure}
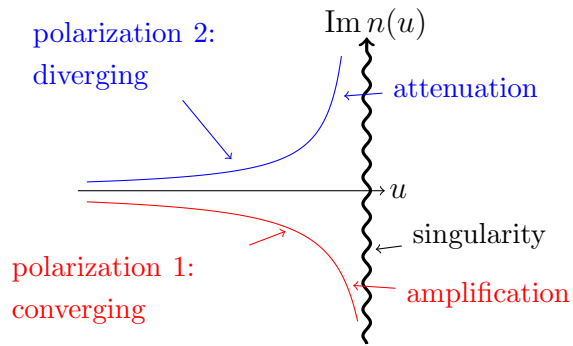

We now follow the photon as it approaches the singularity.  
The amplitude is given by
\begin{equation*}
{|{\cal A}^{(i)}(u)|\over|{\cal A}^{(i)}(u_1)|} = \exp\left[ - \omega \int_{u_1}^u du'~
\IM n_{ii}(u;\omega) \right]\ ,  
\end{equation*}
where we compare with its value at some fixed point $u_1$
in the near-singularity region.  The results are shown for the 
two polarizations in Fig.~\ref{f11}.
As $|u|$ approaches zero, we find
\EQ{
\IM n_i(u;\omega) &= {\alpha\over 12\pi\omega|u|}c_i \ ,
}
with $c_1 = -0.155$ and $c_2 = 0.105$,
which implies that the amplitude itself behaves as a power law:
\begin{equation*}
|{\cal A}^{(i)}(u)|  \thicksim |u|^{\alpha c_i/12\pi}  \ .
\end{equation*}

The fate of the photon is therefore sealed as it falls into the
singularity.  For polarization $\varepsilon^{(2)}$, the curvature 
pulls virtual $e^+e^-$ pairs from the bare field, intensifying the
virtual cloud and screening the bare field so that
its amplitude vanishes.   For the other polarization,
$\varepsilon^{(1)}$, the gravitational tidal forces squeeze the
virtual cloud out of existence, undressing the photon and revealing the 
bare field itself. Its history has come full circle, from the initial
transient dressing and the playing out of renormalization in real time
as it propagates through curved spacetime, to its final destiny as the
vacuum polarization cloud once more vanishes as it enters 
the black hole singularity.

\vspace{0.2cm}
\begin{center}
{\tiny**************}
\end{center}
\vspace{0.2cm}

\noindent This research was supported in part by STFC grants 
ST/G000506/1 and ST/J00040X/1.
We would also like to thank the TH Division, CERN for hospitality
during the course of this work.

\end{document}